\documentclass[a4paper]{article}
\usepackage{fullpage,amssymb,amsfonts,amsmath}
\usepackage{latexsym}
\usepackage{mathrsfs}
\usepackage{jared04}

\newtheorem{hyp}{Hypothesis}
\usepackage{epsfig,color}
\usepackage{graphicx}

\begin{document}

\title{Considerations of a $k=+1$ Varluminopic Cosmology}
\author{Jared M. Maruskin\footnote{Department of Mathematics, The University of Michigan. jmaruski@umich.edu.}}
\date{November 21, 2005}
\maketitle

\begin{abstract}

Every relativistic particle has 4-speed equal to $c$, 
since $g_{\mu \nu} \frac{dx^\mu}{d\tau} \frac{dx^\nu}{d\tau}
= c^2$.  With the choice of $k = +1$ in the FRW metric, the cosmological scale factor 
$a(t)$ has the natural interpretation of the radius of the sphere 
$S^3_a = \{ x \in \mathbb{R}^4 :
(x, x) = a^2\}$.  Thus, a particle at rest in the cosmological frame has 4-speed
equal to $\frac{da}{dt}$.  This leads us to infer that $\dot a = c$, which respresents
a simple kinematic constraint linking the speed of light to the cosmological scale factor.
This drastically changes the 
$k=+1$ picture from a closed deaccelerating universe to an open accelerating universe, 
 settles the horizon problem, 
and provides for a new cosmological model more appealing to our
natural intuition.  In this paper we shall consider ramifications of this model.

\end{abstract}

\tableofcontents

\part{Discussion and Overture} \label{106}

In this part of the paper, we will introduce our hypothesis, 
which will give us a simple kinematic constraint between the evolution
of the scale factor and the speed of light.  We will then consider
various elements of cosmological theory that need modified if one 
is to consider varluminopic cosmological models.  This will set the stage
for our discussion of the dynamics in the upcoming parts of the paper.

\section{Introduction}

In subsection 1, we will review some empirical evidence which points to 
the $k=+1$ FRW Model.  In subsection 2, we will introduce a kinematic
constraint on this model which will change its resulting dynamics.  
In subsection 3, we will give an outline for the organization of this paper.

\subsection{The Appeal of the $k=+1$ Cosmologies}

\subsubsection{Expansion}

Cosmological observation shows us that our universe is expanding.  This
was first discovered by Edwin Hubble in 1929.  As geometry codifies
the interplay between gravitation and matter, this is a natural implication
of general relativity.  It can be mathematically described independent of 
your choice of universe, by including a time-varying scale factor in the
space-time metric.  However, in the $k=+1$ picture, expansion takes a very
natural and elegent form.  The universe takes the form of a three-dimensional
expanding spherical surface in a four dimensional space, the radial component
being time-like.  In this picture, expansion is no more surprising than
observing several dots on a balloon mutually moving farther apart
as the balloon expands.

\subsubsection{The Horizon Problem}

Cosmologist have measured to temperature of the CMBR to be 
2.728 K with anisotropies in temperature on the level of $10^{-5}$.  
The problem is that this is uniform in every direction in the sky.  
In particular, if we look at two points separated by 180${}^o$, we
record the same background temperature.  The radiation from each
of these points has travelled approximately 98\% of the horizon distance,
from the last scattering surface, to reach us.  Therefore, the two points 
where the radiation originitated are seperated by a distance of 
196\% of the horizon distance.  If the universe is 
taken to be flat, we arrive at a fundamental paradox.  How could two
separate points in space-time, not causally connected, be in thermal
equilibrium?  
However, if the universe is spherical, the paradox is resolved.
All points in space are causally connected, as the entire
universe originates from \textit{a single point} in the past.  Thus, by virtue
of the geometry, our model solves the Horizon Problem without the need
of inflation.

\subsubsection{The Simultaneity Problem}

Another cosmological problem is that the standard Big Bang model, when 
set in a $k=0$ flat FRW universe, violates the principle of simultaneity.  
In this model, the Big Bang occurs everywhere simultaneously.  That is,
on a space time diagram, there is a horizontal cutoff somewhere.  Thus 
for every observer in motion relative to the cosmolgical frame (i.e. us),
there are regions of space (on the surface $t = $ now, where $t$ is 
coordinate time) where the big bang is currently going on, and there
are regions of space where the universe has yet to be born.  This problem
is irrelevant in the spherical $k=+1$ FRW universe.

\subsection{A New Kinematic Constraint}

Despite these observations, the $k=+1$ FRW universe has been found unsatisfactory
and inadequate by the cosmology community over the last several decades.  We would
like to introduce a fundamental constraint on the motion which radically changes
the resulting dynamics of the $k=+1$ FRW picture.  Taking the metric
to have signature $(+1, -1, -1, -1)$, it is a basic result of relativity theory
that 
\be
\label{91}
g_{\mu \nu} \frac{dx^\mu}{d\tau} \frac{dx^\nu}{d\tau} = c^2
\ee
For a particle at rest in the cosmological frame, and choosing a comoving
coordinate system, the particle's four velocity must necessarily take 
the form
\be
\label{92}
\left[ \frac{dx^\mu}{d\tau} \right] = \langle c, 0, 0, 0 \rangle
\ee
in order for the condition \eqref{91} to be satisfied.

Now consider the $k=+1$ FRW picture.  In spherical coordinates,
the particle's position is $\langle a, \rho, \theta, \phi \rangle$.  
If the particle is at rest in the cosmological frame, then 
$\dot \rho = \dot \theta = \dot \phi = 0$, and the particle's four-velocity
becomes
\be
\label{93}
\left[ \frac{dx^\mu}{d\tau} \right] = \left\langle \frac{da}{d\tau}, 0, 0, 0 
\right\rangle = \left\langle \frac{da}{dt}, 0, 0, 0 
\right\rangle 
\ee
where $t$ is the proper age of the universe (i.e. cosmologicaly time).  

Equating \eqref{92} with \eqref{93}, we have the following fundamental
cosmological hypothesis:

\begin{hyp} The universe is spherical and the speed of light varies
with cosmological time $t$ subject to the constraint
\be
\label{100}
\dot a(t) = c(t)
\ee
\end{hyp}

Having a speed of light which varies with cosmological time 
is no stranger than having a scale factor which varies with 
cosmological time.

\subsection{Outline}

In Part \ref{106}, we will devote our attention to necessary modifications of 
classical cosmology that should be taken into account when dealing with 
varluminopic theories.  In \S \ref{107}, we will state the metric Ansatz connected
with our hypothesis \eqref{100}.  In \S \ref{108}, we will make a brief comment
on physical constants.  In \S \ref{109}, we will show that an extra term appears
in the geodesic equations when dealing with a varluminopic theory.  This arises
as the geodesic equations are \textit{geometric} and not \textit{dynamical} equations,
and thus must be written with derivatives with respect to arclength $ds$.
In converting to derivatives with respect to proper time, the chain rule
now takes effect, thus modifying the standard geodesic equations.

In Part \ref{111}, we will apply Einstein's Field Equations to our metric
Ansatz, and look at the resulting dynamics.  Thus in \S \ref{112}, we form 
all of the tensor structures related to the metric, such as the Ricci Tensor,
Einstein Tensor, Christoffel Symbols, and Ricci Scalar.  In \S \ref{113} we write
the resulting field equations, which, unlike their isoluminopic FRW counterparts,
do not yield an acceleration equation, rather  pressure \textit{and} density 
equations.  An additional constraint is needed to specify the resulting dynamics.
However, it uniquely establishes the equation of state in cosmological time,
which was not previously the case.  In \S \ref{114} we find the 
resulting covariant divergence of the field equations, which contain additional
terms due to the varying speed of light.  One now has the freedom to constrain the 
total \textit{physical} energy of the universe to be conserved, which was
not previously possible.  We will explore this scenerio, the so-called 
isoergic case, in \S \ref{98}.  Alternatively, we can impose an 
adiabatic condition, tantamount to constraining the covariant divergence
of the stress-energy tensor to vanish.  In this case, one is forced to pick
up an additional dynamic variable, either $G$ or $\Lambda$ or both.  Thus 
one would require an \textit{additional} constraint to determine the resulting 
dynamics.  

In Part \ref{97}, we introduce a new set of Varluminopic Field Equations.  
We start from the hypothesis that the varluminopic and isoluminopic 
gravitational actions coincide.  We then derive the resulting field equations,
in parallel to their classical derivation, while including varluminopic 
effects.  In \S \ref{95}, we show that the standard Lagrangian density can be
recast in a form dependend only on the metric and its first derivative, as is
done classically, even in the varluminopic case.  In \S \ref{96}, we 
derive the resulting field equations.  Finally, in \S \ref{116}, we impose
the additional energy conservation constraint $\nabla_\mu T^\mu_\nu = 0$,
and determine the resulting dynamical evolution of the universe.  We note
that the extra terms which appear in the Varluminopic Field Equations,
as they appear in \S \ref{96}, save us from having to add additional 
dynamical variables and constraints on the motion, as is the case
when one attempts to impose this constraint using Einstein's Classical Field
Equations.

We would additionaly like to state that some work has been done in 
varluminopic cosmoliges (commonly known as VSL or variable-speed-of-light theories).
See Basset et al \cite{bassett1} and the references contained therein,
Barrow \cite{barrow1} and Magueijo \cite{magueijo1}, for instance.
This paper will present a different approach on varluminopic theories.

\section{Change of Variables} \label{107}

The FRW metric with $k=+1$ is equivalent to the form:
\be
\label{01}
ds^2 = c^2 d\tau^2 = c^2 dt^2 - a(t)^2 \{ d\rho^2 + \sin^2(\rho) d\theta^2
+ \sin^2(\rho) \sin^2(\theta) d\phi^2 \}
\ee
the spatial component of which represents the restriction of the standard Euclidean 
metric in $\mathbb{R}^4$ to the three-sphere $S^3_a$.  Our paper diverges from the classical
analysis of this problem by imposing the kinematic constraint
\be
\label{02}
\frac{da}{dt} = c(t)
\ee
We now introduce the change of variables $t \longrightarrow a(t)$ defined by:
\be
\label{03}
a(t) = \int_0^t c(s) \ ds
\ee
Since $c(t) > 0$, $a(t)$ represents a monotone, increasing function of the cosmological
(coordinate) time $t$.  This transformation can therefore be inverted to express:
\be
\label{04}
\tilde c(a) := c(t(a))
\ee
Under this change of coordinates, our metric Ansatz becomes:
\be
\label{05}
ds^2 = \tilde c(a)^2 d\tau^2 = da^2 - a^2 \{ d\rho^2 + \sin^2(\rho) d\theta^2
+ \sin^2(\rho) \sin^2(\theta) d\phi^2 \}
\ee

\section{A Note on Fundamental Physical Constants} \label{108}

In Einstein's derivation of the classical field equations, he supposes that
the Einstein tensor is proportional to the stress-energy tensor of the field,
i.e.
\be
\label{87}
G_{\mu \nu} = \sigma T_{\mu \nu}
\ee
where $\sigma$ works out to be
\be
\sigma = \frac{8\pi G}{c^4}
\ee
where $G$ is Newton's constant, which appears in Newton's Law of Gravitation:
\be
\mathbf{F} = - \frac{GMm}{|\mathbf{r}|^3} \mathbf{r}
\ee

When considering varluminopic theories, we must discover whether it is 
$\sigma$ or $G$ which is to be held constant.  Had Einstein come up with the 
field equations \eqref{87} first, then Newton's Law would have been 
written
\be
\mathbf{F} = - \frac{\sigma c^4 Mm}{8 \pi |\mathbf{r}|^3} \mathbf{r}
\ee
and it would have been $\sigma$ which would have been considered the 
fundamental constant of gravitation.  

In this paper, we will take $G$ to be the fundamental constant.  
We would like to remark that this is not entirely obvious, and is an 
assumption.  There is no distinction in the isoluminopic models, however
it now makes a decided difference when we treat the speed of light as
a dynamic variable.

\section{Modified Geodesic Equations} \label{109}

If one considers the motion of a particle at rest in the cosmological frame (where $ds^2 = da^2$), 
one would
expect that the dynamics of the scale factor should be determined by solving the geodesic
equations associated with the metric \eqref{05}.  One might naively mistake this evolution
to be nontrival, i.e. $\ddot a = 0$.  This conclusion would be invalid.  What is actually
being observed is the obvious condition that $\frac{d^2 a}{da^2} = 0$.  

In our theory; in fact, in any variable speed of light (VSL) theory, one can modify the 
geodesic equations to give \textit{dynamical} equations of motion, as opposed to 
the standard \textit{geometric} equations of motion.  This distinction does not exist
in classical general relativity, where the speed of light is a constant and can be moved
freely through differential operators.

The geometric geodesic equations, which are valid even in the VSL case, are the standard
geodesic equations of motion:
\be
\label{06}
\frac{d^2 x^\mu}{ds^2} + \Gamma^\mu_{\nu \la} \frac{dx^\nu}{ds}\frac{dx^\la}{ds} = 0
\ee
However, using the relation
\be
\label{07}
ds^2 = c(t)^2 d\tau^2
\ee
where $t$ is the local cosmological time, one sees that:
\bea
\label{08}
\frac{dx^\mu}{ds} &=& \frac{dx^\mu}{d\tau} \frac{1}{c(t)} \\
\label{09}
\frac{d^2x^\mu}{ds^2} &=& \frac{d^2x^\mu}{d\tau^2} \frac{1}{c(t)^2} 
- \frac{dx^\mu}{d\tau} \frac{1}{(c(t))^2} \frac{dc(t)}{dt} \frac{dt}{d\tau} \frac{d\tau}{ds} \\
&=& \frac{d^2x^\mu}{d\tau^2} \frac{1}{c(t)^2} 
- \frac{dx^\mu}{d\tau} \frac{1}{(c(t))^3} \frac{dc(t)}{dt} \frac{dt}{d\tau}
\nonumber
\eea
Substituting into the geometric geodesic equation \eqref{06}, one obtains:
\be
\label{10}
\frac{d^2 x^\mu}{d\tau^2} + \Gamma^\mu_{\nu \la} \frac{dx^\nu}{d\tau}\frac{dx^\la}{d\tau} = 
\frac{dx^\mu}{d\tau} \frac{c'(t)}{c(t)} \frac{dt}{d\tau}
\ee
Or, alternatively,
\be
\label{11}
\frac{d^2 x^\mu}{d\tau^2} + \Gamma^\mu_{\nu \la} \frac{dx^\nu}{d\tau}\frac{dx^\la}{d\tau} = 
\frac{dx^\mu}{d\tau} \frac{d\tilde c(a)}{da} \frac{dt}{d\tau}
\ee
We note that if a particle is at rest in the cosmological frame, then $dt = d\tau$.  Substituting
the Christoffel Symbols from \eqref{05} into the modified geodesic equations yield no information
on the dynamics of the scale factor, i.e. they are trivially satisfied.  To determine the dynamics
of the scale factor, we now turn to Einstein's Field Equations.

\section{Redshifts and Distances} \label{110}

We derive the appropriate modifications for redshifts and luminosity 
distances, taking into account the changing speed of light.  See 
Bergstr\"om \cite{bergstrom} and Breton \cite{breton}, among many others,
for a review of the isoluminopic derivation.

\subsection{Redshifts}

Light waves travel along null-geodesics.  Setting the $\theta$ and $\phi$ angular
coordinate displacements to zero, we have, for photons travelling along constant
$\rho$ world lines, the following:
\be
\dot a^2 dt^2 - a^2 d\rho^2 = 0
\ee
For a photon emitted at $t_e$ and absorbed at $t_0$, we thus have
\be
\int_{t_e}^{t_o} \frac{\dot a(t)}{a(t)} \ dt = \int_0^\rho d\rho
\ee
Suppose that the next wave peak is emitted and aborbed $\delta t_e$ and 
$\delta t_o$ later, respectively.  Since they both travel through an angular
displacement $\rho$, we have
\be
\int_{t_e}^{t_o} \frac{\dot a(t)}{a(t)} \ dt
= \int_{t_e + \delta t_e}^{t_o + \delta t_o} \frac{\dot a(t)}{a(t)} \ dt
\ee
From which it becomes clear that
\be
\int_{t_e}^{t_e + \delta t_e} \frac{\dot a(t)}{a(t)} \ dt
= \int_{t_o}^{t_o + \delta t_o} \frac{\dot a(t)}{a(t)} \ dt
\ee
Assuming that $\dot a(t)$ and $a(t)$ are approximately constant over the durations
$\delta t_e$ and $\delta t_o$, we find that:
\be
\label{66}
\frac{\delta t_o}{\delta t_e} = \frac{\dot a(t_e) \la_o}{\dot a(t_0) \la_e} = 
\frac{\dot a(t_e)}{\dot a(t_o)} \frac{a(t_o)}{a(t_e)} 
\ee
where $\la$ is the wavelength.  Therefore the cosmological redshift $z$ is given by
\be
\label{78}
1+z = \frac{\la_0}{\la_e} = \frac{a(t_0)}{a(t_e)}
\ee
which is the regular formula.

\subsection{The Luminosity Distance}

The so-called luminosity distance $d_L$ is measured indirectly by measuring the 
arrival power flux $\mathcal{F}$ of light from distant objects with known
intrinsic luminosities $\mathcal{L}$, by the relation
\be
\label{79}
\mathcal{F} = \frac{\mathcal{L}}{4 \pi d_L^2}
\ee
The luminosity distance is the distance we would be from the object if the universe was
flat and expansionless.  To measure the expansion of the universe, it is 
standard practice to derive a theoretical relationship between the luminosity distance 
and redshifts of incoming light from distant sources, and then compare this relationship 
to observation.  As usual, we must tweek this theoretical relationship to find the 
appropriate one for our new cosmology.

Suppose a photon with energy $E_e$ is emitted from a distant source at time
$t_e$ and with wavelength $\la_e$.  Thus
\be
E_e = \frac{h \dot a(t_e)}{\la_e}
\ee
Its observed energy at arrival will be
\be
E_o = \frac{h \dot a(t_o)}{\la_o} = \frac{\dot a(t_o)}{\dot a(t_e)} \frac{\la_e}{\la_o}
E_e = \frac{\dot a(t_o)}{\dot a(t_e)} \frac{1}{(1+z)} E_e
\ee
The object's intrinsic luminosity $\mathcal{L}$ is the power of a burst of photons
emitted at time $t_e$:
\be
\mathcal{L} = \frac{\delta E_e}{\delta t_e}
\ee
Recalling \eqref{66}, we see that the object's power at arrival is given by
\be
\mathcal{L}_o = \frac{\delta E_o}{\delta t_o} = 
\left[ \frac{\dot a(t_o)}{\dot a(t_e)}\right]^2 \frac{1}{(1+z)^2} \mathcal{L}
\ee
Suppose the constant angular distance to the source is $r$ (where we've replaced
$\rho$ in \eqref{01} to avoid confusion with the energy density).  Then the arrival
flux is given by 
\be
\label{80}
\mathcal{F} = \frac{\mathcal{L}}{4 \pi a(t_o)^2 r^2 (1+z)^2} 
\left[ \frac{\dot a(t_o)}{\dot a(t_e)} \right]^2
\ee
As light travels along a null geodesic from $t_e$ when it is emitted at the 
source to $t_o$ when it arrives at the detectors, we have
\be
0 = \dot a(t) \ dt - a(t) dr
\ee
Hence
\be
r=\int_0^r d \tilde r = \int_{t_e}^{t_o} \frac{\dot a(\tilde t)}{a(\tilde t)} d \tilde t
= \ln \left( \frac{a(t_o)}{a(t_e)} \right)
\ee
This is the simple relation between the angular displacement of a distant object
($a(t_o) r$ being its current instantaneous distance) and the scale factor.
Comparing \eqref{79} to \eqref{80}, we have for the luminosity distance:
\be
\label{81}
d_L = \frac{a(t_0)^2}{a(t_e)} \frac{\dot a(t_e)}{\dot a(t_o)} 
\ln \left( \frac{a(t_o)}{a(t_e)} \right)
\ee
\eqref{78} combined with \eqref{81} provide a theoretical relation between
the redshift and luminosity distance of detectable light.  If one can find 
explicit solutions for $a(t)$ and $\dot a(t)$, one can
then determine the theoretical relationship between $d_L$ and $z$ and compare
with known data.  

In practice, one also computes the distance modulus, defined by
\be
\label{121}
\mu = 5 \log_{10} \left( \frac{d_L}{\mbox{1 Mpc}} \right) + 25
\ee
and plots this measure of distance vs. redshift.


\part{The Classical Field Approach} \label{111}

In this part, we insert our metric Ansatz \eqref{05} into Einstein's Field
Equations and determine the resulting relationships.  For the standard FRW 
metric, the Field Equations produce an equation for the density (the Friedmann
Equation) and an acceleration equation.  Classically, one must 
in addition impose an equation of state to determine the dynamical outcome.
One can gain insight as to how the universe should evolve with various equations
of states, but the equation of state as a function of cosmological time
has not been determined.  

On the other hand, we will show that our metric
Ansatz instead produces a density equation and a pressure equation, 
so that one can completely determine the equation of state as a function
of the dynamic variables $a(t)$ and $c(t)$.  In order to get the 
dynamics, one must pose an additional constraint on the system.  At this 
point we will distinguish two possibilities.  One may take a na\"ive approach
and constrain the total physical energy of the universe to be constant.  
In this case, our so-called ``isoergic'' model, one can determine the 
resulting dynamics of the system.  This approach \textit{cannot} be done
in the isoluminopic models (as the covariant divergence of the Field Equations
vanish), so it has not been seriously considered.  However, its philosophical
implication is that, for it to be correct, there can be no transfer of 
energy from the matter content to the geometry (i.e. gravitational field).
The second, more standard approach, is to set the covariant divergence of the 
stress-energy tensor to zero.  This is the so-called ``adiabatic'' condition.
We will see that as a result of this
condition, one must take either $G$ or $\Lambda$ as an additional dynamical
variable, and thus the dynamical system remains underdetermined.  We would
like to point out that if instead of retaining Einstein's Field Equations 
one retains the gravitational action, a new modified set of 
Varluminopic Field Equations will result, as we will see in Part \ref{97}.
Using the modified Field Equations, the adiabatic condition  will 
completely determine the resulting dynamics without the necessity of 
additional dynamical variables.  The considerations of this
part of the paper are nonetheless interesting; and as the dynamical
solutions in both parts are similar, the work done in this part will provide
insight into the dynamics of the latter.


\section{Einstein's Field Equations} \label{112}

\subsection{The Metric Ansatz}

In this section we state the Christoffel Symbols and nonzero components of the 
Einstein and Ricci Tensors for the metric ansatz stated in \eqref{05}.

\subsubsection*{Christoffel Symbols}
The nonzero Christoffel Symbols are:
\bea
\Gamma^a_{\rho \rho} &=& a \\
\Gamma^a_{\theta \theta} &=& a \sin^2 \rho \\
\Gamma^a_{\phi \phi} &=& a \sin^2 \rho \sin^2 \theta \\
\Gamma^\rho_{a\rho} &=& a^{-1} \\
\Gamma^\rho_{\theta \theta} &=& - \sin \rho \cos \rho \\
\Gamma^\rho_{\phi \phi} &=& - \sin \rho \sin^2 \theta \cos \rho \\
\Gamma^\theta_{a \theta} &=& a^{-1} \\
\Gamma^\theta_{\rho \theta} &=& \cot \rho \\
\Gamma^\theta_{\phi \phi} &=& - \sin \theta \cos \theta \\
\Gamma^\phi_{a \phi} &=& a^{-1} \\
\Gamma^\phi_{\rho \phi} &=& \cot \rho \\
\Gamma^\phi_{\theta \phi} &=& \cot \theta
\eea

\subsubsection*{The Ricci Tensor}
The nonzero components of the Ricci Tensor are:
\bea
R_{\rho \rho} &=& 4 \\
R_{\theta \theta}&=& 4 \sin^2 \rho \\
R_{\phi \phi} &=& 4 \sin^2 \theta \sin^2 \rho
\eea

\subsubsection*{The Ricci Scalar}
The Ricci Scalar becomes:
\be
R = -\frac{12}{a^2}
\ee

\subsubsection*{The Einstein Tensor}
The nonzero components of the Einstein Tensor are:
\bea
G_{aa}&=& \frac{6}{a^2} \\
G_{\rho \rho} &=& - 2 \\
G_{\theta \theta} &=& -2 \sin^2 \rho \\
G_{\phi \phi} &=& -2 \sin^2 \rho \sin^2 \theta
\eea

\subsubsection*{The Stress-Energy Tensor}
As is standard, we will be taking the stress-energy tensor to be that of a perfect
fluid
\be
\label{99}
T_{\mu \nu} = (\ve + P) U_\mu U_\nu - Pg_{\mu \nu}
\ee
where $\ve$ is the energy density, $P$ the pressure, and
\be
U^\mu = \frac{dx^\mu}{ds} = \langle 1, 0, 0, 0 \rangle
\ee
the four-velocity of a particle in the cosmological reference frame.  The stress-energy
tensor can be expressed in matrix form as follows:
\be
[T_{\mu \nu}] = \left( \begin{array}{cccc}
\ve & 0 & 0 & 0 \\
0 & -Pg_{\rho \rho} & 0 & 0 \\
0 & 0 & -Pg_{\theta \theta} & 0 \\
0 & 0 & 0 & -Pg_{\phi \phi} \end{array} \right)
\ee

\subsection{Einstein's Field Equations} \label{113}

Einstein's Field Equations (with cosmological constant $\Lambda$) can be 
written in either of the following two forms:
\bea
\label{12}
G_{\mu \nu} = R_{\mu \nu}- \frac 1 2 R g_{\mu \nu} &=& \frac{8\pi G}{\dot a^4} T_{\mu \nu} + 
\Lambda g_{\mu \nu} \\
\label{13}
R_{\mu \nu} &=& \frac{8\pi G}{\dot a^4} \left( T_{\mu \nu} - \frac 1 2 T g_{\mu \nu} \right) -
\Lambda g_{\mu \nu}
\eea
Einstein's Field Equations in the form of \eqref{12} imply the conditions
\bea
\label{20}
\frac{6}{a^2} &=& \frac{8 \pi G}{\dot a^4} \ve + \Lambda \\
\label{21}
\frac{2}{a^2} &=& \frac{-8 \pi G}{\dot a^4} P + \Lambda
\eea
Alternatively, the Field Equations written as \eqref{13} imply the equivalent conditions
\bea
\label{22}
0 &=& \frac{4\pi G}{\dot a^4} (\ve + 3P) - \Lambda \\
\label{23}
\frac{4}{a^2} &=& \frac{4\pi G}{\dot a^4} ( \ve - P) + \Lambda
\eea
For independent equations of motion,
we will choose the following:
\bea
\label{52}
\ve &=& \frac{\dot a^4}{8 \pi G} \left( \frac{6}{a^2} - \Lambda \right) \\
\label{53}
P &=& \frac{\dot a^4}{8 \pi G} \left( \Lambda - \frac{2}{a^2} \right)
\eea
In \S \ref{98}, we will solve these equations explicitly for the case of a 
matter dominated universe, i.e. assuming $E = \ve V$ is constant.

We can define the following density parameters:
\bea
\label{55}
\Omega_\ve &=& \frac{8\pi G \ve a^2}{3 \dot a^4} \\
\label{56}
\Omega_\Lambda &=& \frac{ \Lambda a^2}{3} \\
\label{57}
\Omega_k &=& -k
\eea
where $k = +1$ in our model, as our metric Ansatz is equivalent to an $k=+1$ FRW 
metric with  the additional kinematic constraint $c(t) dt = da$.  By convention,
$\Omega_\ve$ is the density parameter of the total energy density: matter, radiation,
and any additional vacuum energy besides the cosmological constant.  With these
definitions, \eqref{52} becomes:
\be
\label{58}
\Omega_{tot} = \Omega_\ve + \Omega_\Lambda + \Omega_k = 1
\ee
which is just the regular Friedman's Equation.  
We note that these definitions
are comenserate with the standard ones if you simplify using the kinematic
condition $\dot a = c$.

We would also like to note that $\Omega_\ve = \Omega_R + \Omega_M + \Omega_V$ is taken
to have contributions from radiation, matter, and nonlambdic vacuum energy; obtained
by using $\ve_R$, $\ve_M$, and $\ve_V$ into \eqref{55}, respectively.

\subsection{The Equation of State} \label{101}

Since Einstein's Field Equations determine the energy density $\ve$ and 
pressure $P$ as a function of the scale factor, via \eqref{52} and \eqref{53}, 
they therefore prescribe as well the equation of state.  Introducing the state 
variable $\chi = \Lambda a^2$, we have:
\be
\label{59} 
w(\chi) = \frac P \ve = \frac{\chi - 2}{6 - \chi}
\ee
In particular, we observe
\bea
\label{60}
w(0) &=& - \frac 1 3 \\
\label{61}
w(2) &=& 0 \\
\label{62}
w(3) &=& \frac 1 3 \\
\label{63}
\lim_{\chi \ra 6^-} w(\chi) &=& + \infty \\
\label{64}
\lim_{\chi \ra 6^+} w(\chi) &=& - \infty \\
\label{65}
\lim_{\chi \ra \infty} w(\chi) &=& -1
\eea

\begin{figure}[!h] 
\begin{center}
\includegraphics[width=3in]{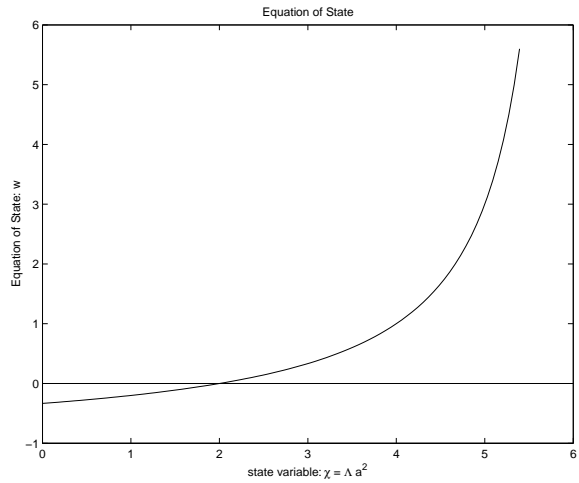}
\end{center}
\caption{Equation of State vs. state variable $\chi = \Lambda a^2$}
\label{123}
\end{figure}

In particular, we note the following important epochs.

\subsubsection{Important Epochs}

\subsubsection*{The Null-$\chi$ Epoch}
When $\chi = 0$, we have the equation of state $P = -\ve/3$.

\subsubsection*{The Dust Epoch}
The universe can be treated as dust for $\chi = 2$.  For $\chi < 2$,
the pressure in the universe is negative.  For $\chi > 2$, 
the pressure is positive.
We are currently very close to this important epoch.

\subsubsection*{The Radiation Epoch}
The universe can be treated as radiation for $\chi = 3$, when its equation
of state is $P = \ve/3$.

\subsubsection*{The Vanishing Epoch}
As $\chi \ra 6^-$, the energy density goes to zero.  At $\chi = 6$,
the universe is completely empty in net energy content.  
After this epoch passes, the universe (if it
continues) shall have negative energy density for all subsequent time.

\subsubsection*{The $\chi$-Infinitum Epoch}
As $\chi \ra \infty$, the universe becomes dominated with a vacuum energy
(unrelated to the cosmological constant term) with an ultimate equation of state
$P = - \ve$.

\subsubsection{$\chi$ and the the Cosmological Constant}

As we are considering a class of cosmologies where the speed of light is
allowed to vary with cosmological time, there is nothing which
prohibits the cosmological constant $\Lambda$ from doing the same.  
For isolambdic cosmologies (with $\dot \Lambda = 0$), 
$\chi$ is an increasing function of cosmological time $t$, since $a(t)$ is.  
This is not necessarily true for their varlambdic counterparts, where evolution
of $\chi$ is determined by $\dot \chi = 2 \Lambda a \dot a + \dot \Lambda a^2$.


\section{The Classical Thermodynamic Analogy} \label{114}

\subsection{Covariant Divergence of the Field Equations}

The Field Equations can be written in the form
\be
\label{67}
G^\mu{}_\nu = \frac{8\pi G}{\dot a^4} T^\mu{}_\nu + \Lambda \delta^\mu_\nu
\ee
where 
\be
[T^\mu{}_\nu] = [g^{\mu \la}T_{\la \nu}] = \left( \begin{array}{cccc}
\ve & 0 & 0 & 0 \\
0 & -P & 0 & 0 \\
0 & 0 & -P & 0 \\
0 & 0 & 0 & -P \end{array} \right)
\ee
is the standard 1-1 stress-energy tensor.

For the moment, we will allow both $G$ and $\Lambda$ to vary with cosmological time,
for the sake of generality.  We only include the case $\dot G \not = 0$ momentarily,
to leave room for future speculations.  Recognizing
\be
\parsh{}{a} =  \frac{dt}{da}\parsh{}{t}
\ee
we obtain for the only nontrivially nonzero component of 
$0 = \nabla_\mu (8\pi G T^\mu{}_\nu/\dot a^4 + \Lambda \delta^\mu_\nu)$
the following:
\be
\label{68}
\frac{8\pi G}{\dot a^4} \left( \frac{\dot \ve}{\dot a} + \frac{3(\ve + P)}{a} \right)
+ \frac{8 \pi \ve}{\dot a} \left( \frac{\dot G}{\dot a^4} - \frac{4 G \ddot a}{\dot a^5}
\right) + \frac{\dot \Lambda}{\dot a} = 0
\ee
This can be compactified into the following equation
\be
\label{69}
\frac{d}{dt} \left( \ln \left( \frac{\dot a^4}{\ve a^3 G} \right) \right)
=  \frac{3w \dot a}{a} + 
\frac{\dot a^4 \dot \Lambda}{8 \pi G \ve} 
\ee
Thus
\be
\label{70}
\frac{\dot a^4}{\ve a^3 G} \propto 
\exp \left( \int \frac{3 w \dot a}{a} + \frac{\dot a^4 \dot \Lambda}{8 \pi G \ve} 
\ dt \right)
\ee

We would like to point out that \eqref{68} imposes no additional constraint on the 
evolutionary dynamics of the system.  By $\ve$ and $P$ satisfying \eqref{52} and 
\eqref{53}, as solutions to the Field Equations, they autmoatically satisfy this
condition.  Whereas the standard FRW metric produces the Frieman Equation and the 
acceleration equation, and leaves open the equation of state; our modification,
when submitted to the Field Equations, determines the equation of state and leaves
open way for an additional constraint which one must impose on the system to 
determine the dynamics.  For example, one now has room  to impose the condition
of isoergiticity on the universe, or adiabaticity, for example, as we will see
in the next section.

\subsection{The First Law of Thermodynamics}

In this section we shall proceed entirely by analogy.  Viewing the universe
as a closed and isolated system, we can compare various terms which arise 
in \eqref{68} to terms which would arise doing a purely classical first 
law control volume analysis.

Allowing $E = \ve V$ be the total energy in the universe, where $V = 2 \pi^2 a^3$
is the total volume of $S^3_a$, we can apply the classical First Law of Thermodynamics
and compare with our conservation condition \eqref{68}:
\be
\dot E = \dot Q - \dot W = \dot Q - p \dot V
\ee
For the given $E = 2 \pi^2 a^3 \ve$ and $V = 2 \pi^2 a^3$, we have
\be
6 \pi^2 a^2 \dot a \ve + 2 \pi^2 a^3 \dot \ve = \dot Q - 6 p \pi^2 a^2 \dot a
\ee
We can meanwhile rewrite \eqref{68} as follows:
\be
\label{71}
6 \pi^2 a^2 \dot a \ve + 2 \pi^2 a^3 \dot \ve = 
\left( \frac{8 \pi^2 a^3 \ve \ddot a}{\dot a} - \frac{\dot \Lambda \dot a^4 a^3 \pi}{4G}
- 2 \pi^2 \ve a^3 \frac{\dot G}{G} \right) - 6p\pi^2 a^2 \dot a
\ee
Thus we can identify 
\be
\label{72}
\dot Q = \frac{8 \pi^2 a^3 \ve \ddot a}{\dot a} 
- \frac{\dot \Lambda \dot a^4 a^3 \pi}{4G}
- 2 \pi^2 \ve a^3 \frac{\dot G}{G} \qquad \qquad \mbox{Cosmological Heating}
\ee
as a Cosmological Heating term, in analogy to classical Thermodynamics.  

In isoluminopic models, we have that $\dot Q \equiv 0$.  And thus it is a 
direct result of the form of Einstein's Field Equations that we must take
$\nabla_\mu T^\mu_\nu = 0$ (which account for the remaining terms of \eqref{71}).
In a varluminopic model, if one uses the classical Field Equations, we are
left with the necessity of imposing an additional constraint, which should
take the form of a conservation law.  We now mention two philosophicaly different
approaches.

\subsubsection{The ``Isoergic'' Condition}

In a so-called isoergic model, we would take $\dot E = 0$.  The varying speed
of light gives us room so that varluminopic effects (encapsulated in the 
$\dot Q$ term) can do the expansion work for us, so that the total 
\textit{physical} energy of the universe is conserved.  In this choice,
physical energy cannot be converted into the gravitational energy of the universe,
as is normally done in the isoluminopic case (though we note that in 
the isoluminopic case, this choice does not exist, and one is forced to 
consider a transfer of physical energy to a gravitational cosmological energy).
The condition that $\dot E = 0$, however, is enough to provide us with 
an acceleration equation, so that the dynamics of the evolving universe can 
be analyzed.  We will do this in \S 98.

\subsubsection{The ``Adiabatic'' Condition}

On the other hand, we do not have to abandon adiabaticity, which
would impose the following dynamical equation of motion
\be
\frac{8 \pi^2 a^3 \ve \ddot a}{\dot a} 
= \frac{\dot \Lambda \dot a^4 a^3 \pi}{4G}
+ 2 \pi^2 \ve a^3 \frac{\dot G}{G} \qquad \qquad \mbox{Adiabatic Condition}
\ee
This is tantamount to imposing the condition that $\nabla_\mu T^\mu_\nu = 0$.
However, it is clear that if one is to obtain nontrivial dynamical solutions
$\ddot a \not = 0$, one must include either $G$ or $\Lambda$ or both as 
dynamical variables.  Thus the system remains underdetermined, and an 
\textit{additional} constraint is needed to determine the dynamics.
Such a constraint can be obtained by fixing the equation of state.  
We will explore this approach in \S \ref{126}.

It is interesting to note that this is not the case with the modified Field
Equations we will derive in Part \ref{97}.  With the Varluminopic Field Equations
one can impose the adiabatic condition $\nabla_\mu T^\mu_\nu = 0$ and completely
determine the resulting dynamics, without the necessity of the introduction
of additional dynamic variables.


\section{Isoergic Models} \label{98}

As mentioned previously, in these new models, we now have the freedom to 
constrain the total energy $E = \ve V$ of the universe to be constant.  
In the isoergic model, the cosmological heating \eqref{72}
is what does the expansion work, so that the net physical energy in the universe is 
conserved.
We will take $\dot G = \dot \Lambda = 0$.
Using these conditions in \eqref{69} leaves us with the following equations of motion
\be
\label{73}
\frac{a \ddot a}{\dot a^2} = \frac{3 w}{4}
\ee
Recalling \eqref{59} and defining $c = \dot a$, we can rewrite this as
\be
\frac{dc}{c} = \frac{-1}{4} \frac{6 - 3 \Lambda a^2}{6a - \Lambda a^3} \ da
\ee
Integrating, we find
\be
\label{75}
\dot a = k (6a - \Lambda a^3)^{-1/4}
\ee
Alternatively, we can proceed directly from the condition $E = \mathrm{const.}$:
\be
8\pi G \ve a^3 = \dot a^4 \left( 6 a - \Lambda a^3 \right) = k^4
\ee
where $k = \sqrt[4]{4 G E_{\rm tot}/\pi}$.
Solving for $\dot a$
we arrive at:
\be
\label{74}
\frac{da}{dt} = \frac{k}{(6a - \Lambda a^3)^{1/4}}
\ee
which agrees with \eqref{75}, but does not require that $G$ and $\Lambda$ be constant.
Assuming $G$ and $\Lambda $ are constants, we can integrate \eqref{74} to obtain
\be
\label{76}
kt = \frac 4 5 \sqrt[4]{6a^5} \ {}_2F_1 \left( \frac{-1}{4}, \frac 5 8 ; \frac {13} 8;
\frac{\Lambda a^2}{6} \right)
\ee
where ${}_2F_1$ is the Gauss Hypergeometric Function (see \cite{crc}):
\be
{}_2F_1(\alpha, \beta; \gamma; \delta) = \sum_{n=0}^\infty \frac{(\alpha)_n 
(\beta)_n}{(\gamma)_n n!} \delta^n
\ee
which has radius of convergence $|\delta| < 1$.  Here, $(a)_n$ is the shifted factorial
\be
(a)_n = \frac{\Gamma(a + n)}{\Gamma(a)}
\ee
and $\Gamma(z)$ is the $\Gamma$ function:
\be
\Gamma(z) = \int_0^\infty t^{z-1} e^{-t} \ dt
\ee
The solution \eqref{76} can be simplified as:
\be
\label{77}
kt = \frac{4 \sqrt[4]{6a^5}}{\Gamma(-1/4)} \sum_{n=0}^\infty
\frac{\Gamma(n - 1/4)}{(5+8n)n!} \left( \frac{\Lambda a^2}{6} \right)^n
\ee

\begin{figure}[!h] 
\begin{center}
\includegraphics[width=3in]{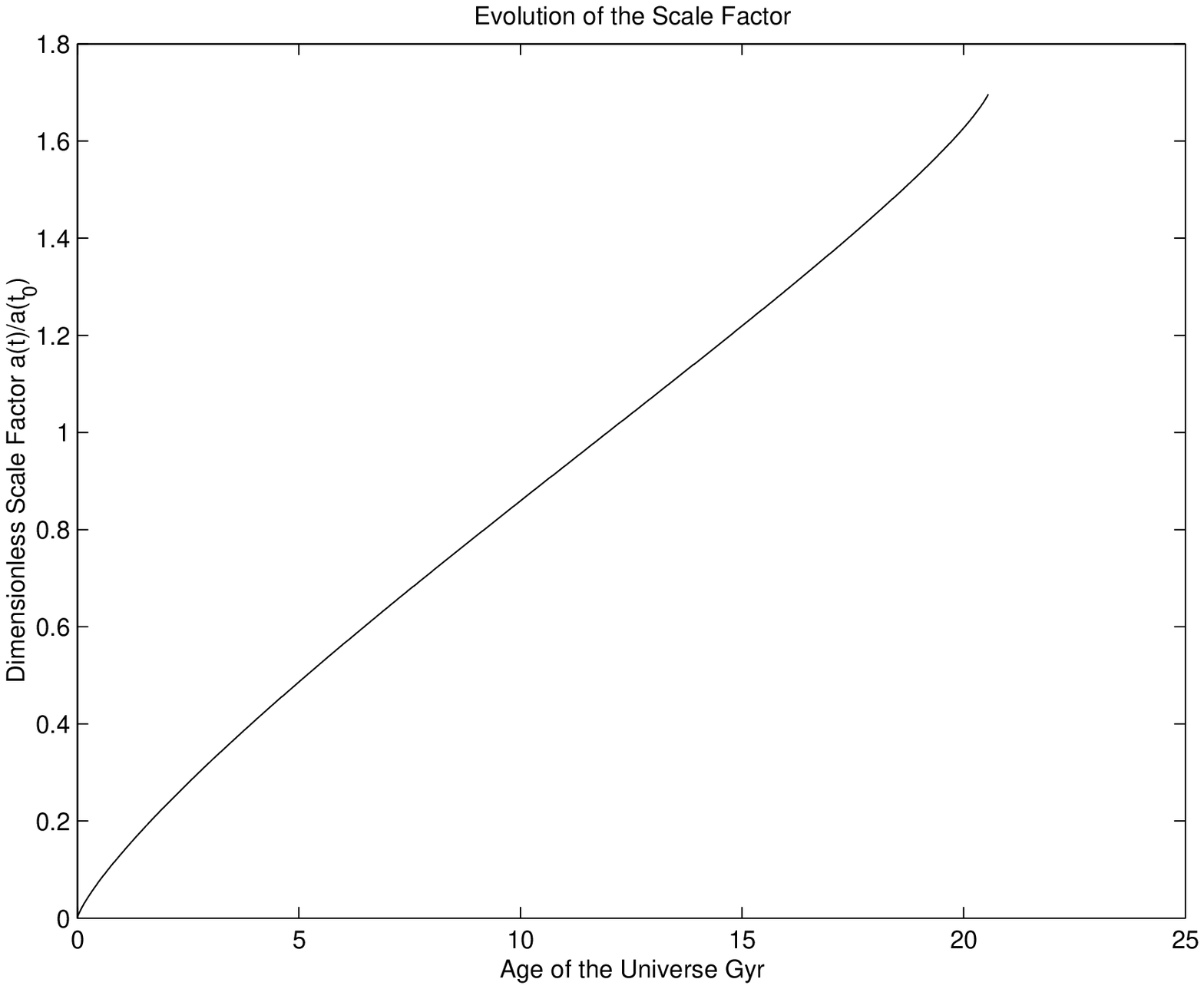}
\includegraphics[width=3in]{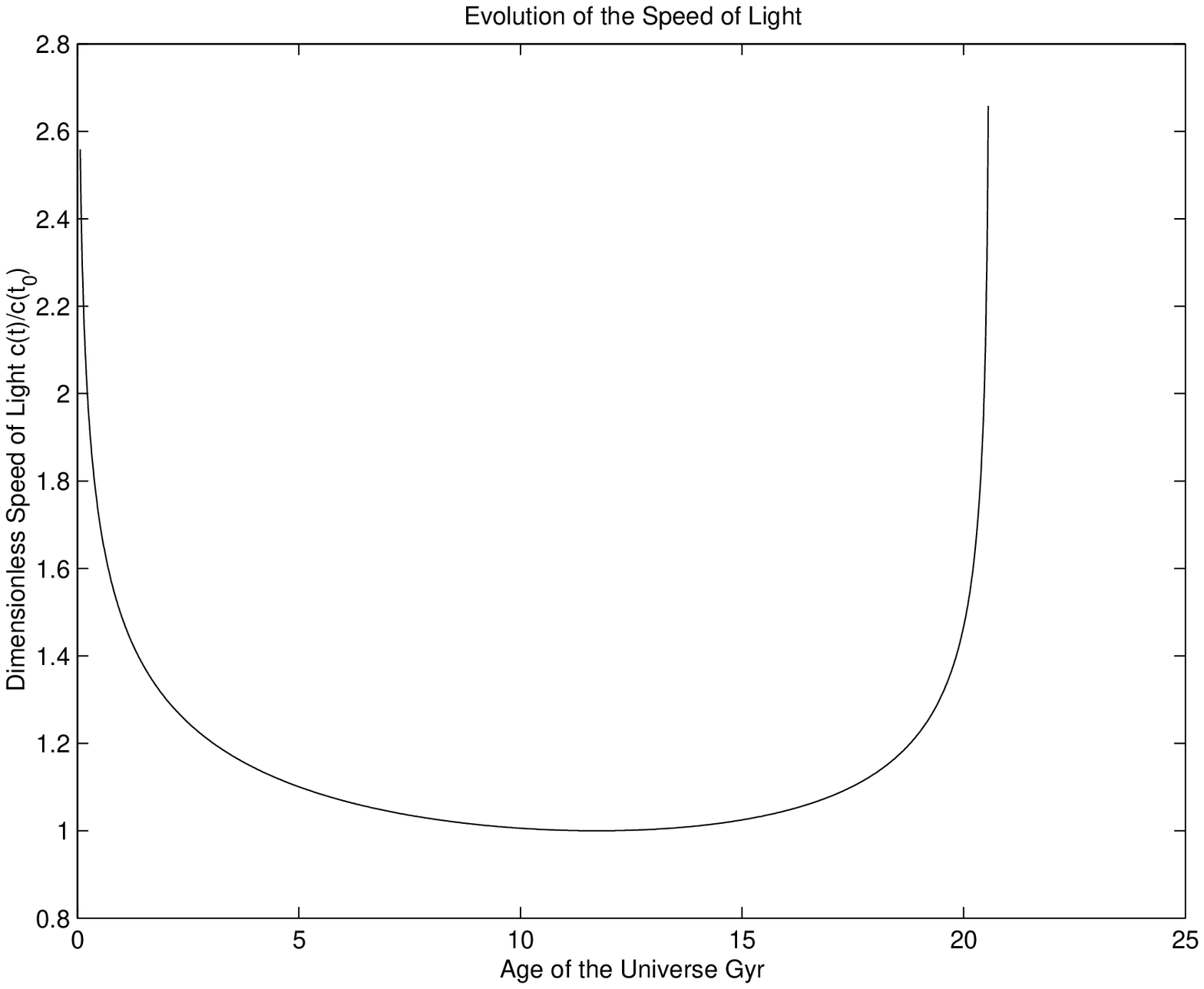}
\end{center}
\caption{$a(t)/a(t_0)$ vs. $t$ (left) and $c(t)/c(t_0)$ vs. t (right)}
\label{124}
\end{figure}

We plot the speed of light \eqref{74} in Fig. \ref{124} left 
and the scale factor in Fig. \ref{124} right.

The radius of convergence of \eqref{77} is $\Lambda a^2 < 6$.  At the epoch
$\Lambda a^2 = 6$, the total energy density of the universe vanishes.
Beyond this point, the energy density would have to become negative for the 
universe to continue, a situation we view as unphysical.  We can use \eqref{76}
with the following mathematical fact (which holds for $c - b - a > 0$, 
as in our case):
\be \label{130}
{}_2F_1(a,b;c;1) = \frac{\Gamma(c) \Gamma(c-b-a)}{\Gamma(c-b) \Gamma(c-a)}
\ee
to see that this model predicts an end to the universe in the finite time
\be
T_{\rm end} = 
\frac{ 4 \cdot 6^{7/8} \Gamma(13/8) \Gamma(5/4)}{5 k \Lambda^{5/8} \Gamma(15/8)}
\approx 20.4 \mbox{ Gyr}
\ee
where we used $a = 1.32 \times 10^26$m and $\Lambda a^2 = 2.1$ (so that
$\Omega_\Lambda = .7$).  These numbers imply the present age of the universe
to be close to 12 Gyr, so that we are currently 8.4 Gyr from the end of the 
universe.

\begin{figure}[!h]
\begin{center}
\includegraphics[width=3in]{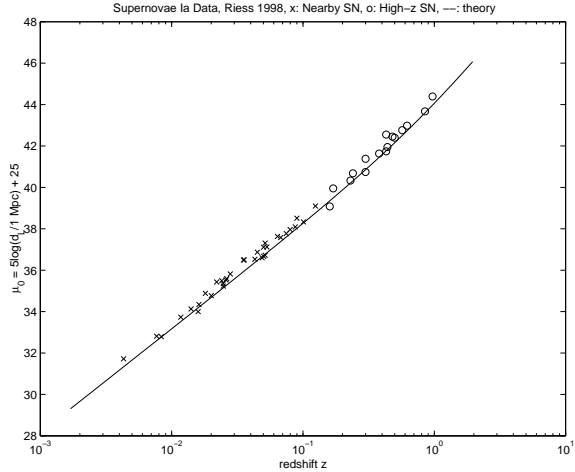}
\end{center}
\caption{Supernovae Ia data vs. theoretical predictions}
\label{125}
\end{figure}

Moreover, using the redshift formula \eqref{78}, the luminosity 
distance formula \eqref{81}, and the definition \eqref{121},
we can compare the theoretical curve produced by this model against 
data points from Supernova Ia data as recorded by Riess \cite{riess}.
We find our theoretical curve matches the supernova quite well, as is
shown in Fig. \ref{125}.

Without the Comsological Constant $\Lambda$, this solution degenerates to
\be
a(t) = \frac{1}{6^{1/5}} \left( \frac{5 k t}{4} \right)^{4/5}  
\ee
Inverting \eqref{77} gives the scale factor, $a(t)$,  as a function of time.


\part{Varluminopic Field Theory} \label{97}

In Isoluminopic Field Theory, Einstein's Field Equations can be derived by 
varying the following action
\be
\label{94} 
S = \int \left\{ \frac{c^4}{16 \pi G} ( \mathfrak{R} + 2 \Lambda \mathfrak{U} )
 - \mathfrak{M} \right\} dt d^3\theta
\ee
Gothicized variables denote tensor densities.  For instance, $\mathfrak{R} = 
R \sqrt{-g}$ is the tensor density of the Ricci scalar, and 
$\mathfrak{U} = \sqrt{-g}$ is the tensor density of unity.  
$\mathfrak{M}$ represents the Lagrangian density of the mass-energy which
occupies the space-time, eg. matter, radiation, etc.  It is well known 
that one cannot construct a scalar density form the metric and its first 
derivatives alone.  Thus the Ricci scalar density represents the most general
covariant candidate for a gravitational action.  Even though it contains 
second derivatives of the metric, these have no effect when varying the 
action \eqref{94}, see Landau and Lifshitz \cite{landau}, Tolman \cite{tolman},
or Pauli \cite{pauli}, amongst others.  Following the derivation in 
Landau and Lifshitz with appropriate modifications, we 
will show in \S \ref{95} that this is true
even when taking the time-varying speed of light into account.

The action for any Varluminopic Field Theory should therefore reduce to 
\eqref{94} in the isoluminopic limit.  The most natural candidate for an 
action is therefore \eqref{94} itself.  In \S \ref{96} we will derive the 
resulting field equations from the action \eqref{94}, taking the varying speed
of light into account.  We will then apply these new field equations to the 
$k=+1$ FRW model, including the kinematic constraint $\dot a = c$, to determine
the dynamical equations of motion of the resulting theory.  We will see
that imposition
of the additional energy conservation constraint $\nabla_\mu T^\mu_\nu = 0$ is 
enough to determine the resulting dynamics explicitly; 
unlike with the classical Field Equations, where one was left
with the necessity of having to also include $G$ or $\Lambda$ as dynamical 
variables.

In formulating the variational principle for gravity, we will find that the 
cosmological coordinates $(t, \rho, \theta, \phi)$ will be better suited 
for our needs.  We will keep these coordinates in the back of our mind while
we derive the modified field equations.  There are two important features 
in doing this.  Notice that all factors of $c$ in \eqref{94} are attached 
solely to the gravitational action.  Thus, by letting $x^0 = t$, as is 
in our prerogative to do so, we 
decouple the varluminopic effects from the matter action.  In standard 
classical derivations of the Field Equations, the entire integrand of \eqref{94}
is typically divided by an extra factor of $c$, in exchange for 
swapping the $dt$ with a $da$ (in the notation allowed by our insight $da = c dt$).
Moreover, in this choice of variables we have $g_{00} = c^2$, so that variations
of $c$ can be easily related to variations of the metric, as will be shown 
in \S \ref{96}.


\section{The Action of the Gravitational Field} \label{95}

As is standard in relativity theory, one identifies the space-time metric
with the graviational field.  It is well-known that there is no scalar
density which depends only on the metric and its first derivatives.  
Thus the Ricci tensor is taken as the key ingredient of the Lagrangian
density of the gravitational field.  Including factors of $c$ which must be
present for our later considerations, the action is similar to the 
nominal action
\[
S_n = \int c^4 \mathfrak{R} dt d^3 \theta
\]
where 
\[
\mathfrak{R} = R \sqrt{-g}
\]
is the Ricci scalar density.  Our aim is to show that this covariant
Lagrangian density is equivalent to one which only involves the metric
and its first derivatives (but depends on the coordinates).  Thus we will show
that
\[
\int c^4 \mathfrak{R} dt d^3\theta = \int \mathfrak{E} dt d^3\theta +
\int \parsh{\mathfrak{w}^i}{x^i} dt d^3\theta
\]
where $\mathfrak{w}^i$ is a vector density, which, by means of Gauss' Theorem,
can be converted to an integral over the boundary and therefore ignored,
and $\mathfrak{E}$ is the 
pseudo-scalar density which depends only on the metric and its first derivative.
We will follow closely the discussion in Landau and Lifschitz \cite{landau}
\S 93, with appropriate modification for the varluminopic effects.

The integrand may be expanded as
\be
c^4 \sqrt{-g} R = c^4 \sqrt{-g} g^{ik}R_{ik}
= c^4 \sqrt{-g} \left\{ g^{ik} \parsh{\Gamma^l_{ik}}{x^l} - 
g^{ik} \parsh{\Gamma^l_{il}}{x^k} + g^{ik} \Gamma^l_{ik} \Gamma^m_{lm}
-g^{ik} \Gamma^m_{il} \Gamma^l_{km} \right\}
\ee
Notice the first two terms may be written as
\bea
c^4 \sqrt{-g} g^{ik} \parsh{\Gamma^l_{ik}}{x^l} &=&
\parsh{}{x^l} \left( c^4 \sqrt{-g} g^{ik} \Gamma^l_{ik} \right)
- \Gamma^l_{ik} \parsh{}{x^l} \left( c^4 \sqrt{-g} g^{ik} \right) \\
c^4 \sqrt{-g} g^{ik} \parsh{\Gamma^l_{il}}{x^k} &=&
\parsh{}{x^k} \left( c^4 \sqrt{-g} g^{ik} \Gamma^l_{il} \right)
- \Gamma^l_{il} \parsh{}{x^k} \left( c^4 \sqrt{-g} g^{ik} \right) 
\eea
The first term on the right hand side of either of these equations may 
be dropped as it may be converted into a boundary integral which vanishes,
as we will take the variations to vanish on the boundary.
Thus we have
\bea
\mathfrak{E} &=& c^4 \left\{ \Gamma^m_{im} \parsh{}{x^k} \left( \sqrt{-g} g^{ik} \right)
- \Gamma^l_{ik} \parsh{}{x^l} \left( \sqrt{-g} g^{ik} \right) -
c^4 \left( \Gamma^m_{il} \Gamma^l_{km} - \Gamma^l_{ik} \Gamma^m_{lm}
\right) g^{ik} \sqrt{-g} \right\} \nonumber \\
& & + 
\Gamma^m_{im} \sqrt{-g} g^{ik} \parsh{(c^4)}{x^k} -
\Gamma^l_{ik} \sqrt{-g} g^{ik} \parsh{(c^4)}{x^l} \\
&=&
c^4 \mathfrak{g}^{ik} (\Gamma^m_{il} \Gamma^l_{km} - \Gamma^l_{ik} \Gamma^m_{lm})
+ \Gamma^m_{im} \mathfrak{g}^{ik} \parsh{(c^4)}{x^k} -
\Gamma^l_{ik} \mathfrak{g}^{ik} \parsh{(c^4)}{x^l} \\
&=&
c^4 \mathfrak{g}^{ik} (\Gamma^m_{il} \Gamma^l_{km} - \Gamma^l_{ik} \Gamma^m_{lm})
+ (\Gamma^m_{im} \mathfrak{g}^{il}  -
\Gamma^l_{ik} \mathfrak{g}^{ik}) \parsh{(c^4)}{x^l}
\eea
This pseudo-scalar density only depends on the metric and its first derivative,
and we have that
\be
\delta \int c^4 \mathfrak{R} dt d^3\theta = 
\delta \int \mathfrak{E} dt d^3 \theta
\ee
Thus we shall not hesitate in using $\mathfrak{R}$ as a Lagrangian density,
even though it contains second derivatives of the metric, as they have
no effect when one takes the variation.

\section{The Varluminopic Field Equations} \label{96}

We wish to find the corresponding equations of motion by varying the 
action \eqref{94} with respect to the space-time metric.  The only difference
between our procedure and the classical approach is that $c$ now varies
with cosmological time.  Hence we need to determine the effects of this on
the resulting equations of motion.  Noting that
\be
\label{131}
\delta \sqrt{-g} = - \frac 1 2 \sqrt{-g} g_{ik} \delta g^{ik}
\ee
we see that the variation of the first part of the integrand of \eqref{94} is
\be
\label{84}
\delta \left( c^4 
 \mathfrak{R} \right) = 
\delta \left( c^4 g^{ik} R_{ik} \sqrt{-g} \right)
= c^4  \left( \mathfrak{R}_{ik} - \frac 1 2 \mathfrak{R} g_{ik} \right)
\delta g^{ik} + c^4 \mathfrak{g}^{ik} \delta R_{ik} + 4 \mathfrak{R} c^3 \delta c
\ee
In order to deal with the $c^4 \mathfrak{g}^{ik} \delta R_{ik} $ term,
we will again closely follow the derivation in Landau and Liftshitz \cite{landau},
with the appropriate modifications necessary for our varluminopic considerations.
Choosing a locally geodesic frame, we have that
\be
c^4 g^{ik} \delta R_{ik} = c^4 g^{ik} \parsh{}{x^l} (\delta \Gamma^l_{ik})
- c^4 g^{il} \parsh{}{x^l} (\delta \Gamma^k_{ik})
\ee
Now consider the vector
\be
w^l = c^4 g^{ik} \delta \Gamma^l_{ik} - c^4 g^{il} \delta \Gamma^k_{ik}
\ee
Taking the divergence with respect to $x^l$, we find
\be
\parsh{w^l}{x^l} = 
c^4 g^{ik} \parsh{}{x^l} (\delta \Gamma^l_{ik})
- c^4 g^{il} \parsh{}{x^l} (\delta \Gamma^k_{ik})
+ \left( g^{ik} \delta \Gamma^l_{ik} - g^{il} \delta \Gamma^k_{ik} \right)
\parsh{(c^4)}{x^l}
\ee
Returning to an arbitrary reference frame, we see that
\be
c^4 g^{ik} \delta R_{ik} = \frac{1}{\sqrt{-g}} \parsh{\mathfrak{w}^l}{x^l}
- \left( g^{ik} \delta \Gamma^l_{ik} - g^{il} \delta \Gamma^k_{ik} \right)
\parsh{(c^4)}{x^l}
\ee
We will now show that the extra term which arises (second term on the 
right hand side) actually vanishes.  Recalling that
\be
\delta g_{lp} = - g_{lk} g_{ps} \delta g^{ks}
\ee
we have that
\bea
\delta \Gamma^k_{ik} &=& \delta \left( g_{ls} g^{ks} \Gamma^l_{ik} \right) \\
&=& g_{ls} \Gamma^l_{ik} \delta g^{ks} + g^{ks} \Gamma^l_{ik}
\delta g_{ls} + g_{ls} g^{ks} \delta \Gamma^l_{ik} \\
&=& g_{ls} \Gamma^l_{ik} \delta g^{ks} + g^{rp} \Gamma^l_{ir}
\delta g_{lp} + g_{ls} g^{ks} \delta \Gamma^l_{ik} \\
&=& g_{ls} \Gamma^l_{ik} \delta g^{ks} -
 g^{rp} \Gamma^l_{ir} g_{lk} g_{ps} \delta g^{ks} 
+ g_{ls} g^{ks} \delta \Gamma^l_{ik} 
\eea
Thus
\be
g^{il} \delta \Gamma^k_{ik} = \Gamma^l_{sk} \delta g^{ks}
- \Gamma^l_{ks} \delta g^{ks} + g^{ki} \delta \Gamma^l_{ik}
\ee
and therefore 
\be
g^{ik} \delta \Gamma^l_{ik} - g^{il} \delta \Gamma^k_{ik} = 0
\ee
so that
\be
\int_\Omega c^4 \mathfrak{g}^{ik} \delta R_{ik} dt d^3\theta = \int_\Omega
\parsh{\mathfrak{w}^l}{x^l} dt d^3 \theta =
\int_{\pa \Omega} \mathfrak{w}^l d^3 x =  0
\ee
as we take all variations to vanish on the boundary.

We are therefore justified to write the variation of the full action 
\eqref{94}
\bea
 & & \delta  \int \left\{ \frac{c^4}{16 \pi G} ( \mathfrak{R} + 2 \Lambda \mathfrak{U})
 - \mathfrak{M} \right\} dt d^3\theta \nonumber \\
& & \label{85} = \int \left[ \left\{ \frac{c^4}{16 \pi G} 
\left( \mathfrak{G}_{ik} -  \Lambda \mathfrak{g}_{ik} \right) - 
\frac{\mathfrak{T}_{ik}}{2}
\right\} \delta g^{ik} + \frac{c^3}{4 \pi G} 
\left( \mathfrak{R} + 2 \Lambda \mathfrak{U} \right) \delta c \right] dt d^3\theta
\eea
where 
\be
\mathfrak{G}_{ik} = \mathfrak{R}_{ik} - \frac 1 2 \mathfrak{R} g_{ik}
\ee
is the Einstein tensor density.

Writing the variation of \eqref{94} in the form \eqref{85}, we are now able
to see the full advantage of considering our variational principle
using the coordinate $x^0 = t$ as cosmological time as opposed to 
$x^0 = a$, as is classically the choice.  First, one no longer has a factor 
of $c^{-1}$ appearing in the matter action, so that the varluminopic effects
limit themselves to effects on the action of the gravitational field.  Moreover,
with the choice of $t$ for our cosmological coordinates, we have the relation
\be
g^{00} = \frac 1 {c^2}
\ee
so that 
\be
\delta g^{00} = - \frac{2}{c^3} \delta c
\ee
or 
\be
\label{132}
\delta c = - \frac{c^3}{2} \delta g^{00}
\ee
and thus the variation of \eqref{94} can be written, for 
our choice of cosmological coordinates, as:
\bea
 & & \delta  \int \left\{ \frac{c^4}{16 \pi G} ( \mathfrak{R} + 2 \Lambda \mathfrak{U})
 - \mathfrak{M} \right\} dt d^3\theta \nonumber \\
& & \label{86} = \int  \left\{ \frac{c^4}{16 \pi G} 
\left( \mathfrak{G}_{ik} -  \Lambda \mathfrak{g}_{ik} \right) - 
\frac{\mathfrak{T}_{ik}}{2} 
- \frac{c^6}{8 \pi G} 
\left( \mathfrak{R} + 2 \Lambda \mathfrak{U} \right) \delta^0_i \delta^0_k
\right\} \delta g^{ik}  dt d^3\theta
\eea
Thus, the modified Field Equations, in our choice of cosmological 
coordinates, become
\be
\label{88}
G_{\mu \nu} = \frac{8 \pi G}{c^4} T_{\mu \nu} + \Lambda g_{\mu \nu} 
+ 2 c^2 (R + 2\Lambda) \delta^0_\mu \delta^0_\nu
\ee
Consider now a general coordinate system $x^\mu$.  To every point in 
space-time we identify a cosmological scale factor $a(x^\mu)$, which 
represents a scalar function on the space-time.  As the map $a(t)$ is 
one-to-one, which identifies the cosmological age of the universe with 
the scale factor, we also have the scalar function $t(x^\mu)$ which identifies
the cosmological age of the universe to every point in the space time.  
Transforming \eqref{88} to an arbitrary choice of coordinates, we therefore find
\be
\label{89} 
G_{\mu \nu} = \frac{8 \pi G}{c^4} T_{\mu \nu} + \Lambda g_{\mu \nu} 
+ 2 c^2 (R + 2\Lambda) \parsh{t}{x^\mu} \parsh{t}{x^\nu}
\ee
Moreover, recalling $da = c(t) dt$, we can alternatively express the extra
term as a tensor product of the divergence of the scale factor $a(x^\mu)$ 
with itself:
\be
\label{90}
G_{\mu \nu} = \frac{8 \pi G}{c^4} T_{\mu \nu} + \Lambda g_{\mu \nu} 
+ 2 (R + 2\Lambda) \parsh{a}{x^\mu} \parsh{a}{x^\nu}
\ee
Hence, even though we choose to work in the cosmological frame to simplify
computations, the modified Field Equations can nonetheless be written in 
an entirely covariant fashion.


\section{A Cosmological Action} 
Suppose we would like to add a cosmological action of the form
\be
S_c = \int \mathfrak{C} dt d^3 \theta
\ee
to the action \eqref{94}, where $\mathfrak{C}$ is of the form
\be
\mathfrak{C} = C(a, c) \mathfrak{U}
\ee
We have
\be
\delta \mathfrak{C} = \parsh{C}{a} \mathfrak{U} \delta a
+ \parsh{C}{c} \mathfrak{U} \delta c + C \delta \mathfrak{U}
\ee
However, noting that
\be
\delta c = \frac{dc}{dt} \frac{dt}{da} \delta a
\ee
we can write $\delta a$ in terms of a corresponding $\delta c$.
Using this relation with \eqref{131} and \eqref{132}, we find
\be
\delta \mathfrak{C} = \left\{
- \frac{c^3}{2} \left( \frac{\dot a}{\dot c} \parsh{C}{a} + \parsh{C}{c}
\right) \delta^0_\mu \delta^0_\nu - \frac{C g_{\mu \nu}}{2} \right\}
\sqrt{-g} \delta g^{\mu \nu}
\ee
Adding this to \eqref{86} and converting to an arbitrary coordinate system,
we obtain the following field equations:
\be
G_{\mu \nu} = \frac{8 \pi G}{c^4} T_{\mu \nu} + \left(\Lambda + \frac C 2 \right)
g_{\mu \nu} + 
\left( 2R + 4 \Lambda + \frac{8\pi G}{c^3} \left( \frac{\dot a}{\dot c} \parsh{C}{a}
+ \parsh{C}{c} \right) \right) \parsh{a}{x^\mu} \parsh{a}{x^\nu}
\ee
However, for the rest of the discussion here we will take $\mathfrak{C} \equiv 0$,
so that our action is to coincide with the classical action for 
general relativity.

\section{Application to our Varluminopic $k=+1$ Model} \label{116}

\begin{figure}[!h] 
\begin{center}
\includegraphics[width=3in]{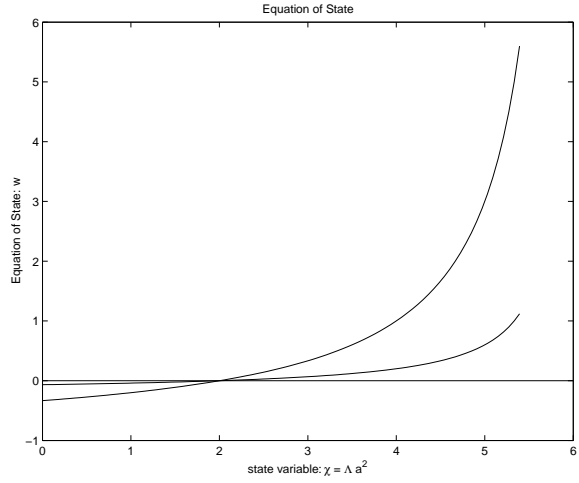}
\end{center}
\caption{Equation of State vs. state variable $\chi = \Lambda a^2$ 
for classical and modified Field Equations} \label{118}
\end{figure}

Using the Varluminopic Field Equations \eqref{90} with our metric Ansatz 
\eqref{05} and the stress-energy tensor for a perfect fluid 
\eqref{99}, coupled with our hypothesis \eqref{100}, 
we obtain the following Field Equations (compare with 
\eqref{20} and \eqref{21}):
\bea
\frac{6}{a^2} &=& \frac{8\pi G}{\dot a^4} \ve + \Lambda - \frac{24}{a^2}
+ 4\Lambda \\
\frac{2}{a^2} &=& - \frac{8\pi G}{\dot a^4} P + \Lambda
\eea
These may alternatively be written in the form (compare with 
\eqref{52} and \eqref{53}):
\bea
\label{104}
\ve &=& \frac{5 \dot a^4}{8 \pi G} \left( \frac{6}{a^2} - \Lambda \right) \\
\label{105}
P &=& \frac{\dot a^4}{8 \pi G} \left( \Lambda - \frac{2}{a^2} \right)
\eea
These have the same qualitative features as those
for the classical Field Equations approach as discussed in \S \ref{101}.
In particular, notice that \eqref{104} implies a new Friedmann Equation:
\be
\Omega_\ve + 5 \Omega_\Lambda + 5 \Omega_k = 1
\ee
Also, the equation of state is now given by 
\be
\label{120}
w(\chi) = \frac P \ve = \frac{\chi - 2}{5(6 - \chi)}
\ee
where $\chi = \Lambda a^2$ (compare with \eqref{59}).  The 
equation of state is plotted in Fig. \ref{118}.  We also included
the equation of state obtained by using the classical field equations
\eqref{59} on the same plot for reference.  The equation of state
for the modified field equations, as given by \eqref{120}, is the 
curve which has been vertically compressed by a factor of 5.

In particular, we note that the matter-dominated epoch occurs
at $\chi = \Lambda a^2 = 2$ and that the radiation-dominated epoch
occurs at $\chi = \Lambda a^2 = 4.5$.  Recall that using the 
classical field equations, as in \S \ref{112}, these epochs occured
at $\chi = 2$ and $\chi = 3$, respectively.

Applying the additional constraint that $\nabla_\mu T^\mu_\nu = 0$,
one can derive (either directly from the stress-energy tensor, or by 
incorporating this condition when taking the covariant divergence of 
the Varluminopic Field Equations \eqref{90}):
\be
\frac{a \ddot a}{\dot a^2} = - \frac 1 5 \frac{6 - 3 \Lambda a^2}{6 - \Lambda a^2}
\ee
Setting $c = \dot a$, we have
\be
\frac 1 c \frac{dc}{da} = - \frac 1 5 \frac{6 - 3 \Lambda a^2}{6a - \Lambda a^3}
\ee
And therefore, integrating, we obtain:
\be \label{102}
\frac{da}{dt} = \frac{k}{ (6a - \Lambda a^3)^{1/5}}
\ee
where $k$ is a constant of integration, \textit{different} than the 
constant of integration which appears in \eqref{74}.  \eqref{102} may 
now be integrated to find
\be
kt = \frac 5 6 \sqrt[5]{6a^6} \ {}_2F_1\left( - \frac 1 5, \frac 3 5 ;
\frac 8 5; \frac{\Lambda a^2}{6} \right)
\ee
This can be simplified to 
\be
\label{103}
kt = \frac{5 \sqrt[5]{6a^6}}{2 \Gamma(-1/5)} 
\sum_{n=0}^\infty \frac{\Gamma(n-1/5)}{(3+5n)n!} \left(
\frac{\Lambda a^2}{6} \right)^n
\ee

\begin{figure}[!h] 
\begin{center}
\includegraphics[width=3in]{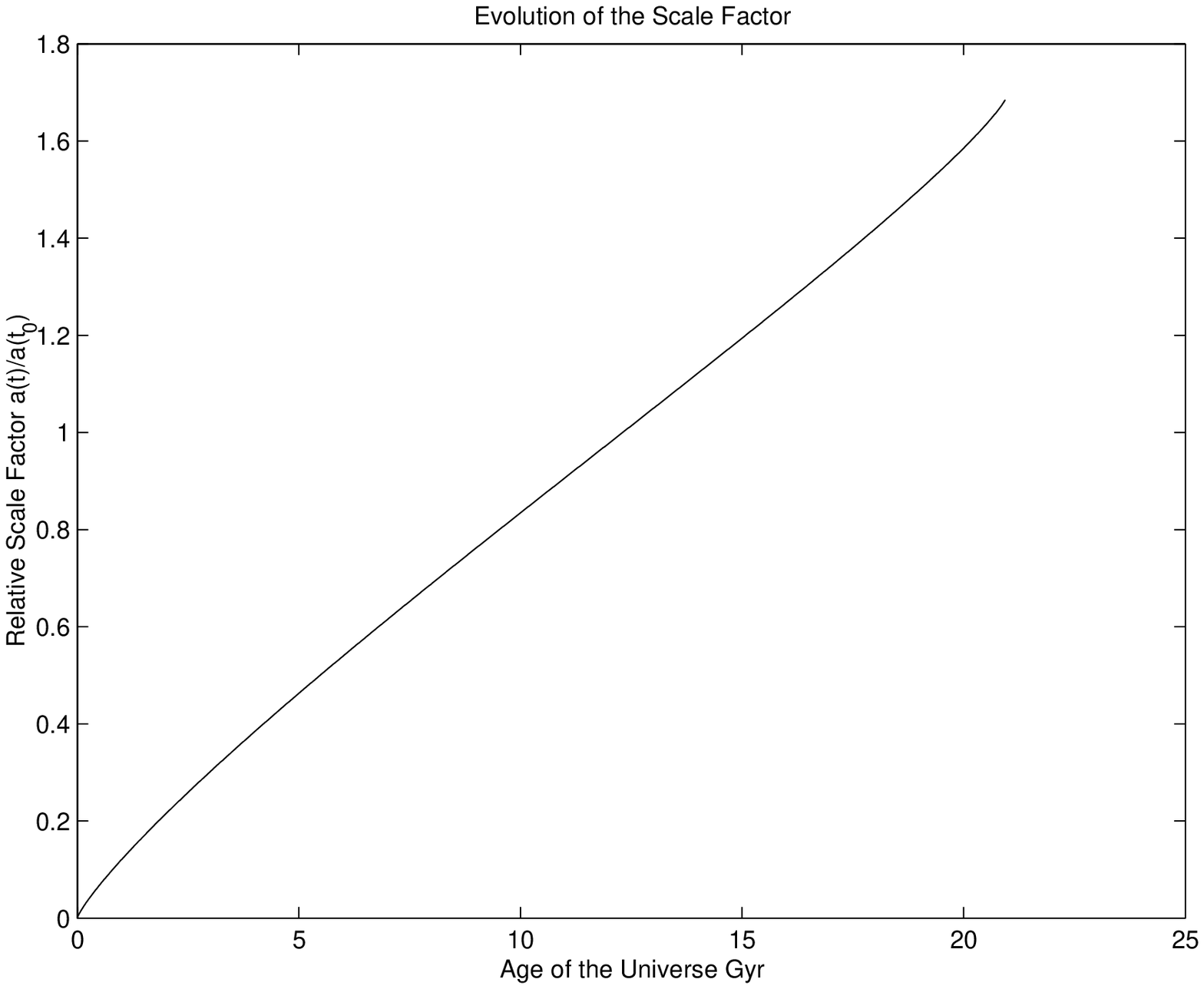}
\includegraphics[width=3in]{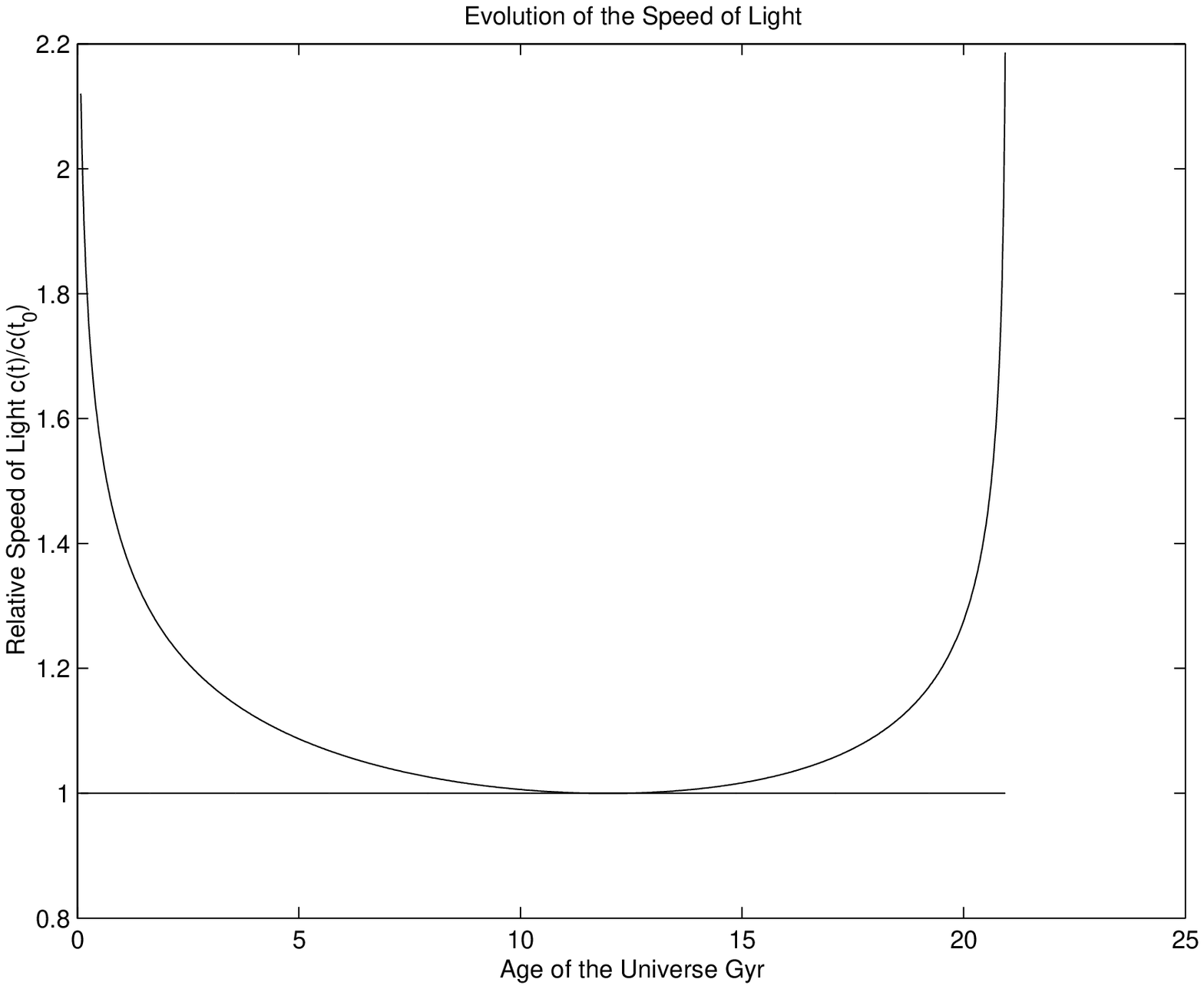}
\end{center}
\caption{$a(t)/a(t_0)$ vs. $t$ (left) and $c(t)/c(t_0)$ vs. t (right)}
\label{119}
\end{figure}

The scale factor \eqref{103} and the speed of light \eqref{102} are plotted
against time in Fig. \ref{119}.  We would like to point out that, despite 
appearences, we are slightly past the minimum value $c_{min}$ in
Fig. \eqref{102}r.  We used $a(t_0) = 1.32 \times 10^{25} \mbox{m}$
and $\Lambda a^2 =2.1$ (so that $\Omega_\Lambda = .7$).

\begin{figure}[!h] 
\begin{center}
\includegraphics[width=3in]{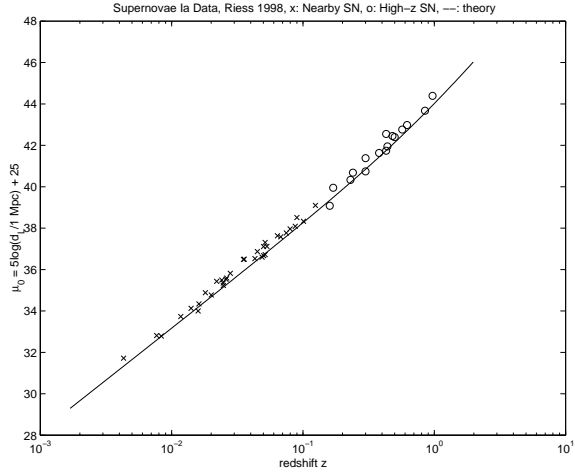}
\end{center}
\caption{Supernova Ia data vs. theoretical prediction} \label{122}
\end{figure}

Moreover, using the redshift formula \eqref{78}, the luminosity 
distance formula \eqref{81}, and the definition \eqref{121},
we can compare the theoretical curve produced by this model against 
data points from Supernova Ia data as recorded by Riess \cite{riess}.
We find our theoretical curve matches the supernova quite well, as is
shown in Fig. \ref{122}.

We would like to note that the scale factor evolution, as seen in 
Fig. \ref{119}, and the theoretical redshift-distance curve, Fig. \ref{122},
are nearly identical in appearance to those predicted by the isoergic model from
\S \ref{98}.  See Fig. \ref{124} and Fig. \ref{125}.

At the epoch $\Lambda a^2 = 6$, we see that the speed of light approaches
infinity.  This epoch also corresponds to the dissapearance of mass-energy,
as is indicated by the density equation \eqref{104}.  We interpret this
as an indication that the universe has ``used up'' the 
total amount of mass-energy in the universe by doing the expansion work.  
This is essentially the death of the universe.  If one were to continue
$\Lambda a^2$ beyond the value 6, one would obtain negative energy 
densities, which we view to be unphysical.
Referring to \eqref{130}, we can 
determine the cosmological time of the 
universe's impending demise to be:
\be
\label{117}
T_{\rm end} = \frac{5 \Gamma(8/5) \Gamma(6/5)}{6^{1/5} k \Lambda^{3/5} 
\Gamma(9/5)} \approx 21.0 \mbox{ Gyr}
\ee
This model predicts that the current age of the universe is 
approximately 12.3 Gyr, so that the universe would have  
about 8.7 Gyr of life left.

Alternatively, without a cosmological constant, the solution to 
\eqref{102} would reduce to
\be
a(t) = \frac{1}{6^{1/6}} \left( \frac{kt}{5} \right)^{5/6}
\ee

\section{A Note About the Varlambdic Approach} \label{126}

By introducing a time-varying cosmological constant, one can
effectively ``control'' the equations of state as given by 
the Classical Field Theory (CFT) and Varluminopic Field Theory
(VFT) approaches.  The radiation- and matter-dominated epochs 
occur at
\bea
\label{128}
\chi_{\rm mat}^{\rm CFT} = \chi_{\rm mat}^{\rm VFT} &=& 2 \\
\chi_{\rm rad}^{\rm CFT} &=& 3 \\
\chi_{\rm rad}^{\rm VFT} &=& 4.5
\label{129}
\eea
where $\chi$ is the state variable $\chi = \Lambda a^2$, see 
\eqref{59} for the CFT case and \eqref{120} for the VFT case.
Replacing $\Lambda$ with $\chi$ as a dynamical variable, we 
can choose a function $\chi(t)$ which is initially constant
with $\chi_{\rm rad}$ and whose value changes to $\chi_{\rm mat}$
during a rapid phase transition.  This corresponds to a rapid
phase transition from a radiation- to matter-dominated universe, with 
time varying cosmological constant.  

The Field Equations can be rewritten as
\bea
\ve_{\rm CFT} &=& \frac{\dot a^4}{8 \pi G a^2} (6 - \chi) \\
\ve_{\rm VFT} &=& \frac{5 \dot a^4}{8 \pi G a^2} (6 - \chi) \\
P &=& \frac{\dot a^4}{8 \pi G a^2}(\chi - 2)
\eea
see \eqref{52}, \eqref{53} (CFT case) and \eqref{104}, \eqref{105} (VFT case).
For both cases, the conservation law $\nabla_\mu T^\mu_\nu = 0$
can be written as
\be \label{127}
\frac{ \dot \ve}{\dot a} + 3( \ve + P) a = 0
\ee
Defining the \textit{acceleration} parameter
\be
Q = \frac{a \ddot a}{\dot a^2}
\ee
we have that \eqref{127} produces
\bea
Q_{\rm CFT} &=& \frac 1 2 + 
\frac{a \dot \chi - 12 \dot a}{4 \dot a (6 - \chi)} \\
Q_{\rm VFT} &=& \frac 1 2 + 
\frac{a \dot \chi - 12 \dot a (7 - \chi)/5}{4 \dot a (6-\chi)}
\eea
Interestingly enough, recalling \eqref{128}-\eqref{129}, we see that 
both approaches lead to the same steady-equation-of-state
dynamics:
\bea
Q_{\rm mat} &=& - \frac 1 4 \\
Q_{\rm rad} &=& - \frac 1 2
\eea
Thus, for either the CFT or VFT approach, the scale factor becomes
\bea
a(t)_{\rm mat} &\propto & t^{4/5} \\
a(t)_{\rm rad} &\propto & t^{2/3}
\eea
for a matter-dominated and radiation-dominated universe, respectively.

\part{Conclusion} 

\section{Conclusion}

In this paper, we presented a new cosmology where the speed of light 
varies with cosmological time subject to a fundamental constraint \eqref{100}.
We explored the implications of this in Part \ref{111}
using the classical field equations and showed that it leads to
a new and interesting dynamics of the scale factor.  In Part \ref{97}
we took a different approach.  Retaining the classical gravitational
action \eqref{94}, we showed that one obtains a new set of Field Equations
\eqref{90}.  With the additional constraint $\nabla_\mu T^\mu_\nu = 0$,
we solved these modified field equations explicitly \eqref{103} and
found that they imply a possible end to the universe
which would occur at the epoch $\Lambda a^2 = 6$, or roughly 9 Gyr from
now.  The model predicts a universe which is currently around 12 Gyr old, 
which does not contradict known timelines.  A key feature of the model is that
it solves the Horizon Problem without the need of inflation.

We showed that, although the redshift formula \eqref{78} remains unaltered,
the formula for the luminosity distance \eqref{81} inherits an extra
factor in the varluminopic case.  Using the dynamical
solutions as presented in \S \ref{98} and \S \ref{116}, we plot the 
redshift-distance curve and showed that in both cases they match with the 
experimental measurements recorded in Riess \cite{riess}, see Fig. \ref{125}
and Fig. \ref{122}.  These results need to be examined more closely, but 
the preliminary figures included in this paper are a good sign.

Big Bang nucleosynthesis, on the other hand, could lead to problems.  
Our model changes the conditions at the time of nucleosynthesis quite
drastically.  When the universe was at the temperature $T = $ 1 MeV,
the speed of light, given by \eqref{102}, was 229 times larger than its
present day value.  Thus it is not immediately obvious whether 
nucleosynthesis will still work in our model, and will considered
in future works.  Other astrophysical issues, such as the
cosmic microwave background and the formation of structure, 
also need to be studied if this model is to remain viable.

Our model changes the history of the early universe.  We no longer have the 
need for inflation.  Phase transitions should work out to be different, eg.
electroweak and quark hadron.  And further it changes the conditions
during baryogenesis, which might also be different.  All of these issues
are uncertain in the standard model, and they may be better or worse here.
This will be the focus of much further research.

\section{Acknowledgement}

I would like to express appreciation to the excellent instruction
in General Relativity, from which I greatly benefited, as provided
by M.J. Duff, J.P. Krisch, and J.A. Smoller.  
I would also like to thank D.J. Scheeres and A.M. Bloch for their 
tremendous support and encouragement, and F.C. Adams 
for his helpful feedback and suggestions.  For our miscellaneous discussions on 
general relativity and cosmology, I am further grateful to G. Tarl\'e, 
L.A. Pando Zayas, K. Freese, and D. Spolyar.



\begin{thebibliography}{99}



\bibitem{barrow1} Barrow, J.D. [2003], ``Unusual Features of Varying Speed of 
Light Cosmologies,'' gr-qc/0211074.

\bibitem{bassett1} Basset, B.A., et al. [2000], ``Geometrodynamics of Variable-Speed-of-Light
Cosmologies,'' Phys. Rev. D 62:103518, astro-ph/0001441.

\bibitem{bergstrom} Bergstr\"om, L. and A. Goobar [1999], \textit{Cosmology and Particle
Astrophysics}, John Wiley and Sons.

\bibitem{breton} Bret\'on, N., J.L. Cervantes-Cota, M. Salgado, [2004], \textit{The 
Early Universe and Observational Cosmology}, Springer.

\bibitem{dubrovin} Dubrovin, B.A., A.T. Fomenko, and S.P. Novikov [1992],  
\textit{Modern Geometry-Methods and Applications:  Part I.  The Geometry of Surfaces,
Transformation Groups, and Fields},  Springer-Verlag.


\bibitem{kolb} Kolb, E.W. and M.S. Turner [1990], \textit{The Early Universe}, 
Westview.

\bibitem{landau} Landau, L.D. and E.M. Lifshitz [1997], \textit{The Classical
Theory of Fields}, Elsevier Science and Technology Books.

\bibitem{magueijo1} Magueijo, J. [2003], 
``New Varying Speed of Light Theories,'' astro-ph/035457.

\bibitem{misner} Misner, C.W., K.S. Thorne, and J.A. Wheeler [1973], \textit{Gravitation},
Freeman.

\bibitem{pais} Pais, A. [1982], \textit{'Subtle is the Lord...' The Science and the Life
of Albert Einstein}, Oxford.

\bibitem{pauli} Pauli, W. [1981], \textit{Theory of Relativity}, Dover.

\bibitem{reif} Reif, F. [1965], \textit{Fundamentals of Statistical and Thermal
Physics}, McGraw-Hill.

\bibitem{riess} Riess, A.G., et al [1998], \textit{Observational Evidence
from Supernovae for an Accelerating Universe and a Cosmological Constant},
astro-ph/9805201.

\bibitem{ryden} Ryden, B. [2003], \textit{Introduction to Cosmology}, Addison Wesley.

\bibitem{schutz} Schutz, B.F. [2000], \textit{A First Course in General Relativity},
Cambridge University Press.

\bibitem{tolman} Tolman, R.C. [1987], \textit{Relativity, Thermodynamics, and Cosmology},
Dover.


\bibitem{weinberg} Weinberg, S. [1972], \textit{Gravition and Cosmology:  Principles
and Applications of the General Theory of Relativity}, John Wiley and Sons.


\bibitem{crc} Zwillinger, D. [1996],  \textit{CRC 
Standard Mathematical Tables and Formulae: 30th Edition}, CRC Press.






\end{thebibliography}
\end{document}